# LIBERAR AL GLOBO TERRÁQUEO

*Alejandro Gangui* [1]

**Resumen:** El globo terráqueo paralelo es un dispositivo antiguo, muy simple e ingenioso que empleado en forma sistemática en las clases de astronomía se convierte en una herramienta didáctica de gran potencialidad. Orientado adecuadamente de acuerdo a la meridiana local, este instrumento permite seguir las sombras en cualquier región de la Tierra que esté iluminada por el Sol, además de ofrecer una visualización clara del terminador, la línea en rápido movimiento que divide el día de la noche en nuestro planeta. Con el conocimiento de las sombras, es posible estimar la latitud de un sitio e inferir la hora solar local en cualquier lugar del hemisferio iluminado del planeta. Además, mediante el empleo del globo terráqueo paralelo se puede comprender de manera simple que existan regiones donde los objetos a veces no proyectan sombras y otras, por el contrario, que a veces se convierten en "países de las sombras largas". En este trabajo, primeramente reseñamos el dispositivo, sus fundamentos básicos de armado y funcionamiento. En la segunda parte, describimos en detalle algunas actividades que facilitan su empleo en el aula y que hemos venido desarrollando en talleres de formación docente en nuestro grupo de investigación.

**Palabras-clave:** modelos concretos; didáctica; astronomía; vivencial; globo terráqueo paralelo.

# SOLTAR O GLOBO TERRESTRE

**Resumo:** O globo terrestre paralelo é um antigo dispositivo, muito simples e criativo que, empregado de forma sistemática nas aulas de astronomía, converte-se em uma ferramenta didática de grande potencialidade. Orientado adequadamente de acordo com o meridiano local, esse instrumento permite acompanhar as sombras em qualquer região da Terra que esteja iluminada pelo Sol, além de oferecer uma visualização clara do terminadouro, a linha em rápido movimiento que divide o dia da noite em nosso planeta. Com o conhecimento das sombras, é possível estimar a latitude de uma localidade e inferir a hora solar local em qualquer lugar do hemisfério iluminado do planeta. Além disso, mediante o emprego do globo terrestre paralelo, pode-se compreender, de maneira simples, que existem regiões onde os objetos às vezes não projetam sombras e outras, pelo contrario, que às vezes se convertem em "países das sombras longas". Neste trabalho, primeiramente, descrevemos o dispositivo, seus fundamentos básicos de construção e funcioamento. Na segunda parte, descrevemos em detalhes algumas atividades que facilitam seu emprego na aula e que temos desevolvido em oficinas de formação docente em nosso grupo de pesquisa.

**Palavras chave:** modelos concretos; didática; astronomia; empirismo; globo terrestre paralelo.

# FREE THE GLOBE

**Abstract:** The parallel globe is an old, very simple and ingenious device that, when systematically employed in astronomy classes, becomes a teaching tool with great potential. Properly oriented according to the local meridian, this instrument allows us to follow the shadows in any region of the Earth that is illuminated by the Sun, as well as offering a clear view of the terminator, the fast-moving grey line that divides the day from the night on our planet. With knowledge of the shadows, it is possible to estimate the latitude of a site and to infer local solar time anywhere in the planet's sunlit hemisphere. Furthermore, by using the parallel globe we may understand simply the existence of regions in which objects sometimes do not cast shadows, and also other regions which, on the contrary, sometimes become "long-shadow" countries. In this work, we first review the device and the basics of its assembly and operation.

---

[1] IAFE/Conicet y Universidad de Buenos Aires, Argentina. <relat@iafe.uba.ar>

In the second part, we describe in detail some activities targeted to facilitate its use in the classroom, which our research group has been developing during teacher training workshops.

**Keywords:** didactics; astronomy; experiential; parallel globe.

## 1. Introducción

¿Qué adulto de hoy en día no recuerda la viñeta de Mafalda en donde la hija intelectual de Quino, al observar su ubicación en un globo terráqueo, se sorprendía al verse "cabeza para abajo"? ¿Qué lector no recuerda a Ernesto Sábato cuando en uno de sus libros y refiriéndose a la visión de la Tierra de la ciencia medieval europea, el autor escribía: "San Isidoro no admitía siquiera la existencia de habitantes en Libia, por la excesiva inclinación del suelo [...] por la misma razón que se negaba la existencia de los antípodas, esos absurdos habitantes con la cabeza para abajo"?

Si bien los investigadores en didáctica han venido trabajando desde hace largos años sobre la gravitación y sus dificultades en la enseñanza (por ejemplo, NUSSBAUM, 1979; 1999) y se han escrito incluso tesis enteras (CAMINO, 2006), los obstáculos en incorporar adecuadamente el concepto de gravedad persisten en todos los niveles de la escolaridad y no están ausentes tampoco en la formación docente.

Aunque hoy nos deje perplejos esa antigua idea de los teólogos medievales, quienes quizás invocaban la posibilidad de "resbalamiento" una vez que los navíos se alejaban más allá de una cierta distancia de Europa (SÁBATO, 1970), no son pocos los estudiantes que en la actualidad evidencian su confusión frente a situaciones que ponen en juego la real comprensión de su ubicación en la Tierra y de los elementos astronómicos que nos ayudan a relacionarnos con el entorno.

Desde hace mucho tiempo la investigación didáctica en astronomía aboga por aunar esfuerzos para lograr la construcción de una visión dual (local o vivencial y planetaria) de los fenómenos naturales observables a simple vista, es decir, sin el empleo de instrumentos ópticos. Según Camino (2011), para aproximarse a una correcta interpretación de lo que se observa desde la posición de cada observador (la posición topocéntrica) es necesario construir esta visión dual. Esto puede lograrse a través del empleo conjunto de instrumentos que faciliten la parte vivencial (la observación del horizonte local (ROS, 2009), el estudio de las sombras mediante el empleo del gnomon recto vertical, la estimación de la meridiana del lugar (CAMINO et al., 2009), etc.) y de aquellos que presenten los mismos fenómenos pero vistos desde una descripción más global, es decir desde el mega-espacio (LANCIANO, 1996). El dispositivo que estudiaremos aquí nos permitirá reflexionar sobre este último aspecto y nos brindará la visión planetaria que buscamos.

En el marco de su inspirada defensa de la descripción vivencial de los fenómenos astronómicos (a veces conocida con el nombre de descripción ptoloméica), Lanciano (1989) afirma que "frecuentemente las experiencias escolares muestran que si se recibe información sin plantear ningún problema, sin formular hipótesis, sin preguntas, probablemente no se llegará a retener y a reelaborar la información recibida". En nuestra experiencia frente a los alumnos hemos podido verificar sus dichos, pues creemos que un buen camino para paliar las deficiencias en la construcción de conocimiento científico es la enseñanza-aprendizaje por investigación, que supone enfrentar a los alumnos con diversas preguntas o situaciones problemáticas. De la misma forma que en la ciencia los conocimientos se elaboran en respuesta a preguntas, este enfoque considera que una enseñanza basada en situaciones problemáticas favorece un aprendizaje significativo (PORLÁN, 1999).

El empleo de representaciones concretas, como el dispositivo que aquí discutiremos, permite que los estudiantes puedan pensar e intervenir sobre el mundo de los fenómenos sobre el cual se está trabajando (su ubicación sobre la superficie de la Tierra, su relación espacio-temporal con otros observadores terrestres, etc.). Los modelos concretos deben ser concebidos como instrumentos útiles para acompañarlos en el proceso de resignificación de sus vivencias astronómicas cotidianas (CAMINO, 2004) y en ese camino el rol del docente es fundamental. Lo que proponemos aquí es que, en ese recorrido junto a sus alumnos, el docente busque la manera de motivarlos a abordar el aprendizaje con el recurso al trabajo con problemas, en especial con aquellos interrogantes que sean significativos para los aprendices. Es esta actividad de investigación del alumno la que se relaciona más fuertemente con el proceso de construcción de conocimientos.

En coherencia con lo dicho, en la segunda parte de este trabajo presentaremos una serie de actividades que esperamos puedan facilitar el abordaje del dispositivo en el contexto de la enseñanza. Esto es, seguiremos una metodología didáctica basada en la investigación o, en palabras de PORLÁN (1999), en "la organización de actividades de enseñanza-aprendizaje en torno al planteamiento y resolución de problemas relacionados con el medio natural, con el objetivo de hacer evolucionar las concepciones espontáneas de los alumnos". En este trabajo usaremos de manera novedosa ese elemento tan común y ubicuo de las escuelas de nuestra región: el globo terráqueo, al que en lo que sigue reorientaremos de manera útil para su empleo en nuestras clases de astronomía, convirtiéndolo en lo que se conoce como un "globo terráqueo paralelo", un recurso didáctico de inmenso potencial.

## 2. Globos (des)orientados

Cuando uno mira un globo terráqueo, lo primero que nota es que la esfera terrestre nunca viene sola; sin un soporte adecuado, el globo (que a veces notaremos GT en lo que sigue) podría rodar sobre la superficie donde lo apoyamos (la mesada de la escuela, por ejemplo) y eventualmente caerse. Por ello, el globo viene sujetado –en general, por sus polos geográficos– por un soporte, como se muestra en la Figura 1.

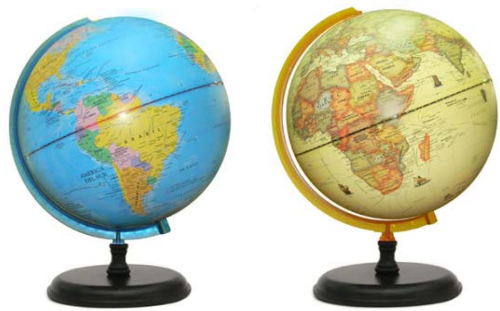

**Figura 1: Los globos terráqueos que se ven en los comercios y en las escuelas, en general, vienen provistos de un soporte que los mantiene "parados". Pero si es cierto que en el espacio no existe ni arriba ni abajo, ¿por qué siempre orientarlos de la misma manera, con el polo norte geográfico hacia arriba? Fuente de numerosas ideas espontáneas entre los alumnos de corta edad ("¿por qué los que viven cerca del polo sur no se caen?"), esta orientación convencional nos obliga a ser muy cuidadosos en el tratamiento de temas sobre la verticalidad y la gravitación como una fuerza hacia**

**el centro, y a observar (y hacer observar) la fuerte dependencia que conceptos como "arriba" y "abajo" tienen de la posición topocéntrica.**

Es aquí que aparece una arbitrariedad. ¿Por qué orientarlo con el hemisferio norte hacia arriba? Por supuesto, una posible respuesta sería: "porque de alguna manera hay que orientarlo". Pero ¿por qué *siempre* se lo ve orientado con el polo norte arriba y el polo sur hacia abajo? (y lo mismo puede preguntarse en el caso de los planisferios). La respuesta más lógica es que esa orientación es simplemente una convención, o sea que no tiene nada de fundamental y que bien podría haberse convenido que el sur quedase arriba y el norte abajo. (En algunos países con geografía "lineal" –y según como se mire, algo vertical– como Italia, estos conceptos están muy arraigados incluso en la manera de hablar. Si un habitante de Trieste decide visitar a un amigo de Palermo, muy probablemente le escriba "vengo giù" –voy para abajo–, sinónimo de "viajo hacia el sur". De la misma manera, al regresar, le dirá "torno su" –vuelvo hacia arriba– para darle a entender que viaja de regreso hacia su ciudad.)

De más está decir que esta convención, que simplifica la fabricación de globos terráqueos (pues todos los nombres de países, ríos, continentes, océanos, etc., están orientados apropiadamente), también da lugar a una inmensa cantidad de ideas espontáneas (muchas científicamente equivocadas) y genera obstáculos en la enseñanza de la Tierra como cuerpo cósmico y sobre la verticalidad (NUSSBAUM, 1999).

Con la idea de reflexionar sobre este instrumento, hace algunos años surgió un proyecto internacional, liderado por la Prof. Nicoletta Lanciano, llamado "el movimiento de liberación del globo terráqueo", que centralizó los esfuerzos de muchos investigadores y docentes interesados en las potencialidades del GT (LANCIANO et al., 2011). Con manifiestas intenciones educativas y culturales, la idea de *liberación* se refiere, en particular, al soporte del GT, y por lo tanto implica libertad para que cada uno de nosotros oriente el globo terrestre como más le guste. Claro que si queremos poder usarlo en la enseñanza de la astronomía, deberemos orientarlo de la manera más conveniente –o más fiel, podríamos decir– de acuerdo a nuestra ubicación en la Tierra (de acuerdo a nuestra posición topocéntrica), y en lo que sigue veremos cómo hacerlo.

Por el momento, imaginemos que estamos en medio del campo y con horizontes despejados. La ciudad más cercana es Santa Rosa de la Pampa (Argentina), pero se halla a muchos kilómetros de nosotros. Pensemos, ¿hacia dónde queda la ciudad inglesa de Londres? Si señalamos con la mano, muy probablemente apuntemos hacia el noreste, y hacia esa dirección debería volar un avión que conectase las dos ciudades. Pero sabemos bien que esa no es la distancia más corta: si pudiésemos construir un túnel por el interior de la Tierra, la distancia más corta sería una recta que, hundiéndose en La Pampa y pasando por debajo del Océano Atlántico, volviese a la superficie en Londres. En otras palabras, ir hacia Londres por el camino más corto no es ir hacia la derecha o la izquierda, sino ir (aproximadamente) hacia "abajo", como se grafica en la Figura 2.

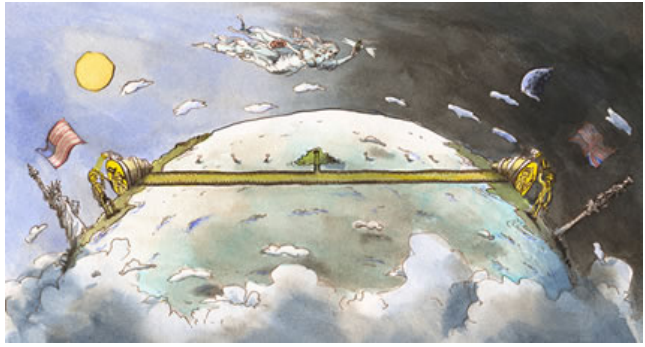

**Figura 2: La distancia más corta entre dos ciudades distantes de la Tierra no es aquella que las conecta por encima de la superficie terrestre, sino aquella que las conecta *atravesando* el globo terrestre. En este caso, las ciudades de Nueva York (hacia la izquierda de la imagen) y de Londres se conectan por un túnel recto que atraviesa el océano atlántico norte. Lo mismo valdría para las ciudades capitales de la provincia de La Pampa y de Inglaterra, mencionadas en el texto.**

Reflexionando un poco, toda ciudad de la Tierra nos queda debajo de nuestros pies. Dado que la superficie terrestre ya no es más un impedimento para nuestro viaje ficticio, pues podemos perforar la tierra como lo haría una lombriz, viajar a cualquier lugar distante es viajar inicialmente hacia abajo de nosotros. De esto concluimos que nosotros (y nuestra ciudad) estamos arriba de todas las demás ciudades de la Tierra. Sabemos además que este "arriba" es un concepto *local*, pues todos los habitantes de nuestro planeta pueden decir lo mismo que nosotros. En efecto, considerando que la Tierra es una esfera perfecta, todos los habitantes del planeta están arriba de todos los demás.

En estas reflexiones se basa el dispositivo que discutiremos aquí, el GT *reorientado*. Pero antes de concentrarnos en él, conviene recordar cómo hallar la meridiana de nuestro lugar de observación (es decir, la dirección norte-sur), pues la precisaremos para alinear adecuadamente el globo y poder hacer las observaciones que deseamos.

## 3. Obtención de la meridiana

Para orientarse ¿qué mejor que una brújula? El empleo de este ingenioso instrumento es sin duda una tarea rápida (si tenemos práctica), bastante precisa (si estamos lejos de yacimientos de hierro, zonas volcánicas y sitios con grandes alteraciones magnéticas), y conveniente (si tuvimos la precaución de colocarla en nuestra valija). Aun así, no aconsejamos usarla, al menos en nuestras clases introductorias de astronomía escolar.

La brújula permite ubicar la dirección del polo norte magnético, no la del polo norte geográfico o astronómico. Para latitudes lejanas de los polos la diferencia no es tan relevante, siempre que no haya alteraciones magnéticas en nuestro sitio de observación. Pero el verdadero motivo por el cual no se aconseja su uso es más metodológico que práctico: constituye un método no astronómico de orientación. Hay además ciertas razones didácticas que hacen a la brújula poco recomendable, pues su empleo fortalece ideas previas sobre la supuesta relación entre el campo magnético terrestre y la rotación del planeta. Además, es muy probable que nuestros alumnos encuentren información sobre varios planetas (no solo de nuestro sistema solar) en los cuales los polos magnéticos se ubican muy lejos de los geográficos. En estos casos la brújula no sería de mucha ayuda para orientarse.

En su trabajo frente a la clase, el docente puede preguntar a los alumnos cómo harían para ubicar el norte sin usar una brújula. Sin duda surgirán muchas ideas, y quizá las sombras ocupen algún lugar entre estas propuestas (GANGUI et al., 2009). Durante un día despejado y sin nubes podemos fácilmente reconocer los cambios de la posición del Sol en la bóveda del cielo. Con ayuda de un *gnomon*, advertiremos que las sombras que arrojan este y todos los objetos expuestos a la luz solar van modificando su longitud y su dirección. Durante las primeras horas posteriores al amanecer, se proyectan sombras largas, siendo cada vez más cortas cuando el Sol se acerca a su punto de culminación. Puesto que el recorrido aparente del Sol es simétrico con respecto a su ubicación en el mediodía solar, encontraremos, a lo largo del día, pares de sombras que tienen la misma longitud. Solo una de ellas no tiene par y es única, la que corresponde al mediodía solar, y que además es la más corta de todas. Esta sombra señala la dirección de la meridiana del lugar, es decir, marca exactamente la línea norte-sur.

Una manera sencilla de obtener la meridiana de lugar consiste en registrar las sombras que proyecta un gnomon sobre el piso durante varias horas de un día cualquiera, especialmente durante las horas cercanas al mediodía solar. En la cercanía de ese momento, cuando la altura del Sol es máxima, será conveniente realizar marcas en el piso cada cinco minutos de manera de reducir lo máximo posible el error observacional (Figura 3).

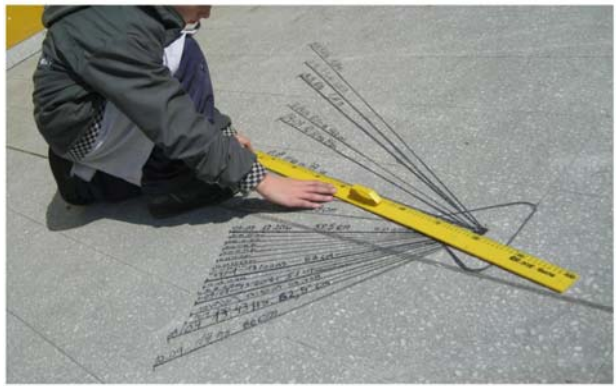

**Figura 3: Medición de las sombras a lo largo del día para obtener la meridiana del lugar. Las sombras arrojadas por un gnomon recto y vertical (no presente en la fotografía) fueron marcadas en el patio de la escuela, consignando para cada una de ellas la fecha y la hora. Con el correr de las horas, las sombras cambian gradualmente de longitud y de dirección; la más corta de todas corresponde a aquella sombra que proyecta el gnomon en el mediodía solar, e indica la dirección norte-sur astronómica verdadera (la imagen es cortesía de M. Iglesias).**

Si no se dispone de tiempo para registrar todo el conjunto de sombras, existe otra forma equivalente y sencilla –incluso, creemos, más precisa– de determinar la dirección norte-sur. Esta nueva manera de usar las sombras para hallar los puntos cardinales consiste en dibujar una circunferencia sobre un papel, o sobre el patio de la escuela, y colocar en su centro un gnomon vertical. Si marcamos con una tiza o un lápiz sobre el piso todos los puntos por los que va pasando la sombra de la punta del gnomon, obtendremos una curva –en nuestras latitudes, por lo común una hipérbola– que cortará a la circunferencia en solo dos puntos. Como podemos imaginar, el primero de estos puntos corresponderá a un momento de la mañana; el otro, a algún momento de la tarde. Y ambos puntos serán equidistantes del mediodía solar. Si ahora unimos con una recta estos dos puntos, tendremos –con muy buena aproximación– la dirección este-oeste. La perpendicular a ésta indicará la dirección norte-sur buscada, como muestra la Figura 4.

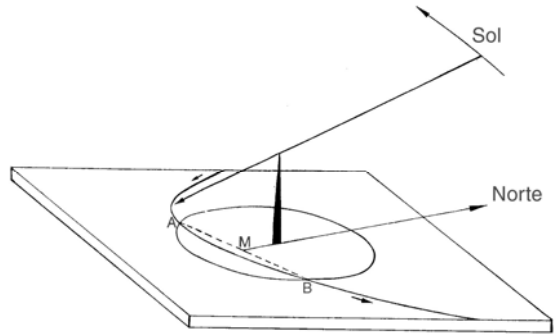

**Figura 4: Método de las sombras para obtener la meridiana del lugar. El esquema representa una situación típica para un sitio del hemisferio sur, donde el Sol jamás llega a pasar por el cenit del observador (por ejemplo, en algún lugar de Sudamérica por debajo del Trópico de Capricornio). Si marcamos todos los puntos del piso donde el extremo del gnomon hace sombra, la curva generada cortará a la circunferencia en dos puntos solamente: el punto "A" a la mañana y el "B" a la tarde, y ambos serán equidistantes del mediodía solar. El punto M (ubicado sobre la recta que une los puntos A y B, y a mitad de camino entre ellos), unido con la base del gnomon, marca la dirección norte-sur. Algo análogo puede hacerse en cualquier sitio de la Tierra (fuera de los polos) donde haya sombras (gráfico adaptado de ROHR, 1986).**

Este método es conocido como el del "círculo hindú", y es conveniente dibujar en el piso no uno, sino varios "círculos", por varios motivos: primero, para asegurarnos de que la sombra del gnomon corte a al menos un par de ellos; segundo, al tener varios puntos "M" (uno por cada circunferencia cortada por la sombra), podemos tomar el promedio de las medidas y reducir así el error de medición (CAMINO et al., 2009).

Este método requiere ser cuidadoso con la elección del radio de la circunferencia. Si la circunferencia es demasiado pequeña, la sombra podría caer por fuera y no cortarla; si es demasiado grande, la sombra siempre la cortará, pero a horarios muy alejados del mediodía solar (y habrá que permanecer muchas horas al Sol). Una opción alternativa es conocer aproximadamente la hora del mediodía solar y efectuar la primera marca ("A" en la Figura 4) unas dos o tres horas antes del mediodía. El tiempo que media entre el momento de la marca "A" y el mediodía solar, es aproximadamente el que deberemos dejar pasar después del mediodía solar para efectuar la marca "B".

Ahora que ya sabemos cómo hallar la meridiana de nuestro sitio de observación, estamos en condiciones de usar el globo terráqueo (GT) que presentamos unas páginas atrás, y convertirlo en un globo terráqueo paralelo.

## 4. Orientando el globo terráqueo

Como veremos, el globo terráqueo paralelo (que con frecuencia notaremos GTP en lo que sigue) es un dispositivo muy simple que, orientado adecuadamente de acuerdo a la meridiana local, nos permitirá hacer un seguimiento de las sombras que proyectan gnomones ubicados en cualquier región de la Tierra (donde sea de día, por supuesto). Por ejemplo, veremos que colocando un palillo sobre la capital de Francia, podremos seguir (en vivo) la sombra que hace la Torre Eiffel sin movernos de Buenos Aires. O bien, con un poco de suerte y colocando un palillo sobre Dubai, la ciudad más poblada de los Emiratos Árabes Unidos, podremos seguir el devenir de la sombra del rascacielos Burj Khalifa, de unos 818 metros de altura, sin desplazarnos de nuestra ciudad. Y quien conoce las sombras de semejantes gnomones, conoce también la hora del día de aquellos lugares y muchas cosas más. Esto es así porque si vemos que nuestra sombra del palillo de Dubai cae al occidente de la meridiana de dicha ciudad árabe, eso significa que la hora solar de Dubai es anterior al mediodía solar, pues el Sol en esa ciudad estará al oriente del meridiano celeste. (Recordemos que el meridiano celeste es la proyección del meridiano terrestre sobre la bóveda del cielo.)

El GTP también nos permitirá materializar muy simplemente el *terminador* en la Tierra y averiguar minuto a minuto en qué comarcas del planeta está amaneciendo y en cuáles se acaba de ocultar el Sol (Figura 5). Con el conocimiento de las sombras sobre el GTP, uno puede además comprender y visualizar ciertas regiones del planeta donde los objetos a veces no proyectan sombras y la existencia de lugares que, por el contrario, a veces se convierten en "países de las sombras largas".

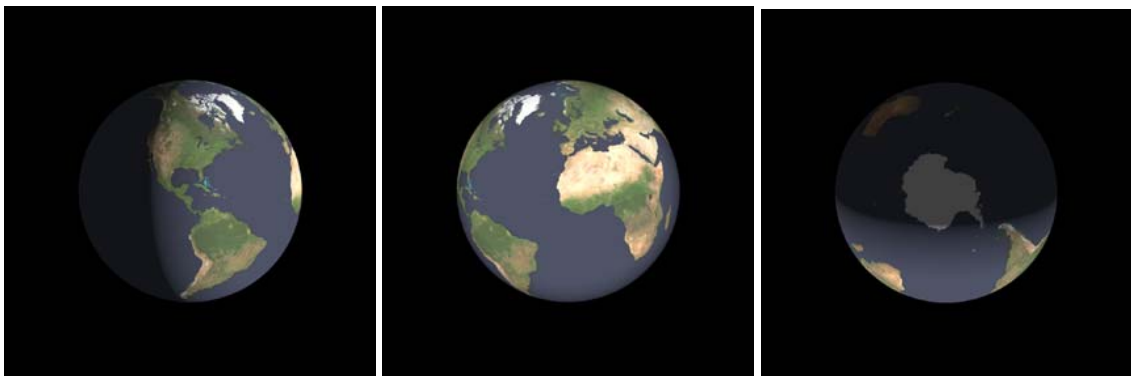

**Figura 5: El terminador de la Tierra se ve claramente en la imagen de la izquierda y en la de la derecha, pero no en la del medio. Las tres son simulaciones de la iluminación de nuestro planeta correspondientes al día 31 de julio de 2013 a la hora 12:52 UT (las 9:52 hora local de la Argentina). La imagen izquierda está vista desde la Luna, es decir, equivale a la fotografía que un habitante de nuestro satélite natural obtendría de la Tierra en ese instante. La de la derecha está vista desde el polo sur celeste y tiene como centro al polo sur de la Tierra (en cuyas cercanías, en esta época del año, jamás ven el Sol). La imagen del medio, que no presenta terminador distinguible corresponde, evidentemente, a una vista desde el Sol (pues "el Sol jamás ve las sombras que él mismo produce"). Imágenes generadas a través de las aplicaciones astronómicas del sitio www.usno.navy.mil/USNO/.**

Como vimos antes, cuando un observador está parado sobre un sitio cualquiera de la superficie terrestre (en la aproximación de la Tierra como una esfera perfecta), todo el planeta queda "debajo de sus pies". Lo mismo sucede si apoyamos el GT en el piso y colocamos un muñequito encima del globo; todos los demás lugares del GT quedarán "debajo" de él. La coincidencia no es casual. Lo que estamos proponiendo aquí es que ambos personajes, el observador real (nosotros, por ejemplo) y el muñequito, nos ubicamos en sitios "semejantes" de nuestros respectivos planetas (la Tierra, en nuestro caso, y el GT en el caso del muñequito).

En este ejemplo, la palabra "semejante" tiene un sentido muy claro: si uno amplifica las dimensiones del GT en un factor aproximado de 4 mil millones (lo suficiente como para pasar de unos 15 cm del radio del GT a unos 6400 km del radio terrestre), nuestro GT se convierte en el planeta Tierra. Y por lo tanto, ese muñequito se convierte en una persona como nosotros. (Ya vemos que el muñequito real debería ser casi invisible para que esta transformación fuera fiel.)

Ahora bien, sabemos que el Sol está tan lejos de la Tierra como lo está de nuestro GT. Para verlo basta comparar los aproximadamente 150 millones de kilómetros del radio de la órbita terrestre ya sea con los 6400 km del radio terrestre o con los 15 cm del radio del globo. El tamaño de la Tierra es casi imperceptible desde esa distancia, igual que lo es el tamaño del GT apoyado sobre nuestro suelo en la superficie del planeta. Es claro entonces que, desde esa distancia, los rayos solares llegan prácticamente paralelos entre sí. E iluminan de igual manera al planeta y al GT, como lo muestra la Figura 6.

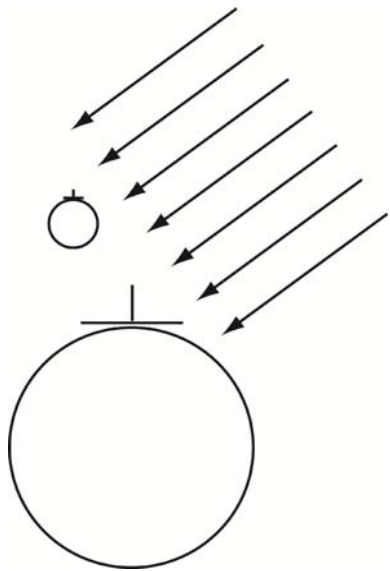

Figura 6: Dos globos semejantes, la Tierra (la esfera grande) y el GT (la pequeña). Ambos globos tienen un gnomon sobre su superficie (un obelisco inmenso en la Tierra y un palillo sobre el GT). Como consecuencia de la gran distancia que nos separa del Sol, los rayos solares (mostrados en el esquema) llegan prácticamente paralelos, tanto sobre el GT como sobre nuestro planeta. En ambos globos se agregaron sendos planos tangentes a la superficie donde se ubican los gnomones. Del esquema se ve claramente que las sombras de los gnomones sobre sus respectivos planos tienen la misma orientación; cuando uno de ellos no recibe luz, el otro tampoco; etc. Las sombras del gnomon menor nos permiten saber cómo evolucionan las sombras del mayor. Nótese que no estamos respetando las dimensiones reales: el globo mayor debería ser unas 4 mil millones de veces más grande que el menor.

Vemos entonces que lo que le pase a la sombra de un palillo colocado sobre el GT será análogo a lo que le pase a una torre inmensamente alta ubicada sobre la Tierra real. Es en este sentido que mirar las sombras sobre el GT nos permite saber lo que sucede en diferentes ciudades de la Tierra, con el solo requisito de que en esas ciudades –y en la nuestra– debe ser de día. (Entre Buenos Aires y Dubai, por ejemplo, hay más de 110° de longitud de diferencia; solo pocas horas separan nuestro amanecer de su puesta de Sol.)

### 5. El globo terráqueo paralelo

Hasta ahora hemos visto que ambos globos (el GT y nuestro planeta) son semejantes, pero eso no basta para nuestra actividad astronómica. Aun debemos hacer dos cosas: primero, debemos elegir bien la ciudad del GT que quedará arriba de todo, y después deberemos orientar el GT adecuadamente. Solo entonces lo que llamamos GT se convertirá en un GTP.

Si queremos usar el GT en nuestra ciudad, nuestra ciudad deberá ir arriba de todo, como lo muestra la Figura 7. Eso es simple de comprender, pues colocándola arriba en el GT todas las demás ciudades del globo quedarán por debajo de ella, como vimos que sucede efectivamente con la ciudad real en el planeta real.

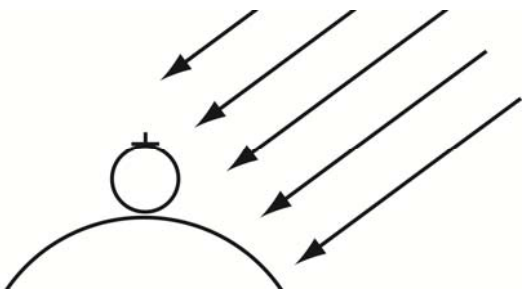

**Figura 7: Los dos globos terrestres, uno ubicado sobre el otro, como sería el caso cuando apoyamos nuestro globo terráqueo escolar a nuestros pies en el patio de la escuela. En este caso, nuestra ciudad debe ubicarse arriba de todo, en el lugar del GT donde hemos colocado el gnomon. Mirando la evolución de las sombras del gnomon pequeño sobre el GT, podemos conocer la evolución de la sombra del mástil de una escuela cualquiera de nuestra ciudad.**

Nos falta ahora fijar un último grado de libertad, y es hacer que las direcciones norte-sur de ambos globos coincidan. En otras palabras, vemos los meridianos de la Tierra dibujados sobre la superficie del GT. En particular, podemos ubicar el meridiano que pasa por nuestra ciudad. También vimos antes cómo dibujar en el suelo la meridiana de nuestro sitio de observación con la ayuda de un gnomon. La idea ahora es hacer que la dirección del meridiano del GT de nuestra ciudad coincida con la dirección de la línea meridiana que marcamos sobre el patio de la escuela. Basta entonces con rotar el GT (dejando nuestra ciudad fija arriba de todo) hasta que ambas líneas sean *paralelas*. Una vez hecho esto habremos alineado ambos globos y podremos afirmar que nuestro globo terráqueo se ha convertido en un globo terráqueo paralelo (Figura 8).

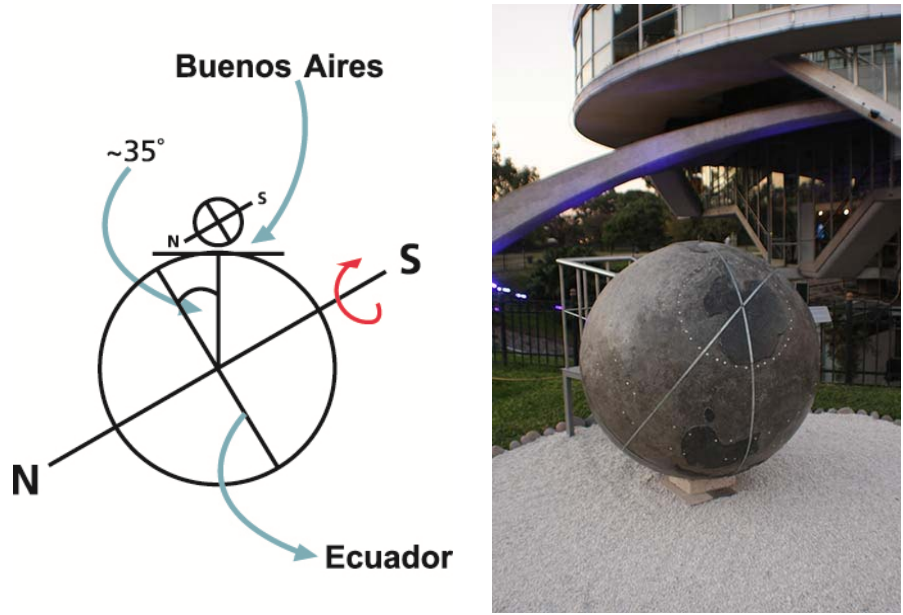

**Figura 8: Para un observador ubicado en el espacio, fuera de la Tierra, ambos ejes de rotación (el del GTP y el del planeta) apuntan en la misma dirección de la bóveda celeste, hacia los dos polos celestes. En la imagen de la izquierda se muestra un esquema del dispositivo alineado para la ciudad de Buenos Aires, ubicada en una latitud aproximada de 35º Sur. Es por eso que dicha ciudad se halla arriba de todo. La fotografía de la derecha muestra el GTP de 1,5m de diámetro construido en la plaza astronómica del Planetario Galileo Galilei de la misma ciudad. Se ve claramente la región Antártica y el polo sur de la Tierra, además de los meridianos de Greenwich (que aparece inclinado) y de la ciudad de Buenos Aires (dispuesto verticalmente, como debe ser).**

¿Qué significa aquí la palabra "paralelo"? Nos indica que los dos ejes de los globos son paralelos en el espacio tridimensional. Es decir, ambos ejes, el de la Tierra y el del GTP, apuntan hacia la misma dirección del espacio. En el hemisferio norte, ambos ejes apuntan hacia el polo norte celeste, que se halla muy cerca de la estrella *Polaris*. En el hemisferio sur, ambos ejes apuntan hacia el polo sur celeste. Globos "paralelos" también significa que el plano tangente en cada punto particular de la superficie de uno de ellos es paralelo al plano tangente del punto correspondiente de la superficie del otro.

Así dispuestos, uno fijo arriba del otro (el GTP apoyado sobre el planeta Tierra), ambos globos rotan alrededor de sus respectivos ejes con el correr de las horas del día y sus diferentes partes reciben la luz solar como ya estamos acostumbrados a ver. No es difícil imaginar que nuestro planeta rota, aunque quizás sea más novedoso pensar que el GTP *también* rota sobre su eje con la misma velocidad angular que lo hace el planeta: ¿cómo lo hace si se halla fijo sobre el piso? Para verlo, basta imaginar una revolución completa de la Tierra arrastrando al GTP y se comprenderá simplemente que ambos globos cumplen una rotación sobre sus ejes (con respecto al Sol) en el mismo tiempo, es decir, en un día de 24 horas en promedio.

## 6. Posibilidades didácticas: terminadores y meridianos

El globo terráqueo paralelo es muy útil para tener una visión global –desde el mega-espacio– de la manera en la que se halla iluminada la Tierra, y permite ver cómo es la evolución del día y de la noche para regiones cercanas o arbitrariamente distantes de nuestra ciudad. Veremos incluso que, prestando atención a la orientación geográfica y a la ubicación del terminador, podremos comprender la ocurrencia de las diferentes estaciones del año para cada sitio sobre la superficie terrestre.

Para iniciar la actividad con el GTP (por ejemplo, en el patio de la escuela y en un día de Sol), luego de haber trabajado junto a nuestros alumnos el tema de su instalación y orientación adecuadas, podemos plantearles, como situación problemática, la siguiente pregunta:

*¿Qué ciudades conocidas están en sombras y cuáles son iluminadas por la luz del Sol?*

El docente puede sugerir que los alumnos anoten esas listas de sitios en sus cuadernos, quizás ordenándolos por cercanía a nuestra ciudad o bien por diferencia de huso horario, pues serán útiles para comparar luego, una hora más tarde, por ejemplo, cuando el Sol haya cambiado un poco su posición.

Se les puede sugerir después que miren la línea del terminador, tanto su "parte" oriental como la occidental. ¿Qué representa esa línea? De un lado, el Sol no llega a iluminar; del otro ya es de día. Discutamos con toda la clase, ¿es realmente una línea perfecta, o parece más bien una alineación de zonas difusas sin un contorno bien definido? ¿Tendrá algo que ver con esto el hecho de que los amaneceres y los atardeceres no son instantáneos, sino que tienen una cierta duración? (un poco más avanzado, ¿tiene algo que ver la atmósfera de la Tierra en todo esto?)

Si nos hallamos en una escuela de Sudamérica, y es de mañana, podríamos discutir con nuestros alumnos sobre la hora que ellos estiman que será en alguna ciudad de Europa. Allá, en el viejo continente, ¿ya es de tarde? ¿Faltará mucho para que se ponga el Sol en Madrid, por ejemplo? Muy probablemente, alguno de los estudiantes recuerde que cuando en Europa central la gente ya terminó de cenar, quienes viven en la ciudad de Rosario, Argentina, apenas están tomando la merienda. De aquí podrán deducir que el terminador oriental sobre nuestro GTP, que inicialmente se hallaba al oriente de Madrid, con el paso de las horas se correrá para el otro lado, hacia el occidente de la capital española, dejándola en las sombras.

Ya vemos entonces de qué manera se mueven las diferentes partes del terminador de la Tierra. ¿Qué ciudades de la lista confeccionada antes están ahora a punto de presenciar un atardecer? ¿Qué ciudades o sitios de la Tierra están ubicados justo *en medio* de las partes occidental y oriental del terminador? ¿Estos representan tan solo algunos sitios o más bien toda una "línea de sitios" sobre la superficie de nuestro planeta? Y si es una línea, ¿coincide con alguna línea ya dibujada sobre el GTP? (Aquí es conveniente mirar en detalle la Figura 9.)

Entre preguntas y discusión, el docente puede llevar a sus alumnos a verificar que la "línea de sitios" en cuestión no es otra cosa que (parte de) un meridiano del globo terráqueo. Y además, podrán verificar que todos esos sitios comparten la misma hora del mediodía solar. (Todas las ciudades que comparten un meridiano tienen siempre la misma hora solar. En este caso particular, esa hora es la del mediodía solar.)

Sin embargo, como es simple de ver con el GTP iluminado por el Sol, la línea (en realidad, la circunferencia) del terminador en general no está representada en el globo; en particular notaremos que en general *no* coincide con uno de los llamados *círculos máximos* de la Tierra, formado en este caso por un meridiano y su antimeridiano (el que se halla separado en 180º de longitud, o sea del otro lado del planeta). Aunque la parte occidental de la línea del terminador nos indica que todas las ciudades que cruza tienen amaneceres en el mismo instante, esas ciudades en general no tienen la misma hora solar, pues como vimos la hora viene fijada por los meridianos.

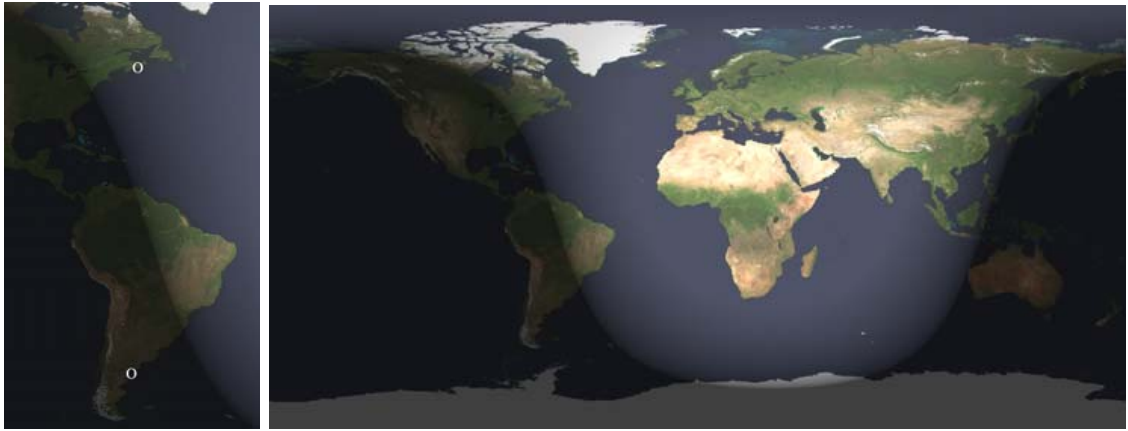

**Figura 9: En la imagen de la izquierda, una persona ubicada en Nueva Escocia (longitud aproximada: 63º O), en el norte de Canadá, y otra en Puerto Madryn (longitud aproximada: 65º O), en el sur de la Argentina, tienen casi la misma coordenada de longitud terrestre y por ello podríamos decir que comparten el mismo meridiano. Sin embargo, en esta imagen del 31 de julio (de 2013) a la hora 10 UT (las 7 AM hora local Argentina), hay casi dos horas de diferencia entre los respectivos amaneceres. En la imagen de la derecha se muestra la proyección del mapa completo de toda la Tierra. De estas imágenes puede comprobarse que el terminador en general no coincide con un meridiano del planeta. Sin embargo, por la simetría evidente en la segunda imagen, vemos claramente que la "línea" (vertical) equidistante entre las partes occidental y oriental del terminador sí coincide con un meridiano terrestre, y es justamente la línea del mediodía solar. Imágenes generadas con software del sitio www.usno.navy.mil/USNO/.**

Por otra parte, es interesante emplear el GTP en un equinoccio, pues entonces se podrá comprobar simplemente que durante esos dos días del año la circunferencia del terminador de nuestro planeta coincide precisamente con la unión de un meridiano y de su antimeridiano. Pues, como el Sol se halla en el ecuador celeste, y por lo tanto pasará por el cénit de los habitantes del círculo ecuatorial, la simetría espacial es completa: durante ese día todo sitio de la Tierra recibirá la luz del Sol y, segundo a segundo, el hemisferio iluminado será limitado por un meridiano y su antimeridiano. Durante un equinoccio es la única oportunidad en la que un observador terrestre y sus antípodas ven la luz del Sol casi simultáneamente.

Sugerimos que el docente enfatice y haga ver a sus alumnos esta diferencia fundamental entre los meridianos y el terminador, pues es el origen de múltiples confusiones en las clases de astronomía, no solo al momento de discutir las diferentes estaciones del año. En particular, el docente podrá hacer ver que, aun compartiendo un meridiano y teniendo la misma hora solar (y el mismo huso horario), dos personas muy alejadas físicamente pueden presenciar amaneceres que distan horas uno del otro (Figura 9).

Ahora bien, ¿qué podemos decir sobre las duraciones del día y de la noche? Usando el dispositivo en fechas bien apartadas de los equinoccios, podemos ver la diferencia en la duración de las horas de luz de distintas partes de la Tierra. Pues algunos paralelos marcados en el GTP tendrán mayor proporción de su longitud iluminada que otros, y eso significa que los habitantes de esos paralelos (de esas latitudes) tendrán mayor cantidad de horas de luz que los habitantes de latitudes muy alejadas (Figura 10).

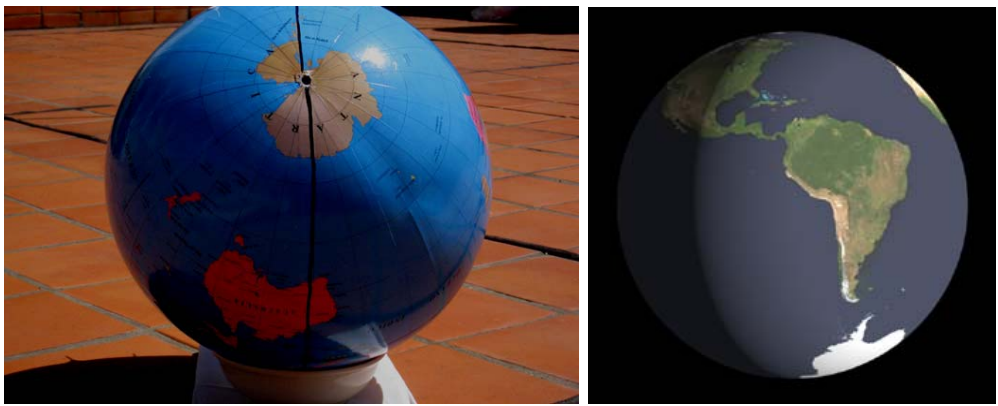

**Figura 10: A la izquierda mostramos un GTP ubicado en la ciudad de Buenos Aires e iluminado por el Sol del 20 de diciembre de 2011, a las 10:30 hora local. El globo está adecuadamente alineado con la meridiana del lugar y la ciudad de Buenos Aires se halla en la parte superior; la fotografía fue tomada desde el sur. A la derecha, una vista diferente de la misma situación, pero a partir de una simulación que nos permite ver a la Tierra desde el espacio o, en el decir de Lanciano (1996), desde el mega-espacio. Ambas imágenes son equivalentes y las regiones iluminadas de ambos globos son idénticas. Se ve claramente que algunos paralelos de la Tierra tienen mayor parte de su longitud iluminada (especialmente aquellos del hemisferio austral) que otros (especialmente del hemisferio norte). Esto es un reflejo de que la situación corresponde al solsticio de diciembre, en el que los días son más largos en el sur que en el norte de nuestro planeta.**

En resumen, sabemos que la Tierra rota sobre su eje y que todos los meridianos efectúan una revolución completa cada aproximadamente 24 horas. Un sitio cualquiera fuera de los polos también cumplirá una revolución en ese tiempo, y lo hará viajando a lo largo de su paralelo (es decir, con coordenada de latitud fija). Si para un día determinado vemos en nuestro GTP que la línea del paralelo que pasa por ese sitio está proporcionalmente poco iluminada (medida en grados de longitud, por ejemplo), la duración de la noche en esa latitud será muy larga y los habitantes de esas comarcas tendrán pocas horas de luz solar. En la situación de la Figura 10, eso les sucede a los habitantes de las regiones del norte, especialmente de aquellas ubicadas hacia el norte del Trópico de Cáncer. En la misma figura vemos que toda la región interior al círculo polar antártico tiene días sin noches, ya que los paralelos de la zona antártica, por ejemplo, están completamente iluminados por el Sol. Como sabemos, esta situación, cuando el Sol pasa días sin ocultarse por el horizonte local, es la razón de lo que se conoce con el nombre de Sol de medianoche.

### 7. La sombra del obelisco

Con alumnos más grandes, además de llevar a cabo la actividad recién detallada sobre los terminadores de la Tierra, podemos comenzar un trabajo sistemático de seguimiento de sombras, como antes lo hicimos con aquellas proyectadas por un gnomon en el patio de la escuela. Esta vez, sin embargo, el o los gnomones deberán ser pequeños y ubicarse sobre la superficie esférica de nuestro globo terráqueo.

Para dar inicio al trabajo recomendamos plantear a los alumnos alguna situación de interés que los convoque con la actividad, que les permita asumir "la problemática a trabajar como un auténtico objeto de estudio" (PORLÁN, 1999). Por ejemplo, en talleres de capacitación que nuestro grupo llevó adelante con docentes (GANGUI e IGLESIAS, 2009), optamos por comenzar haciendo esta pregunta:

*La sombra del Obelisco de la ciudad de Buenos Aires, en este momento, ¿de qué lado de la avenida 9 de Julio se encuentra?* (Figura 11)

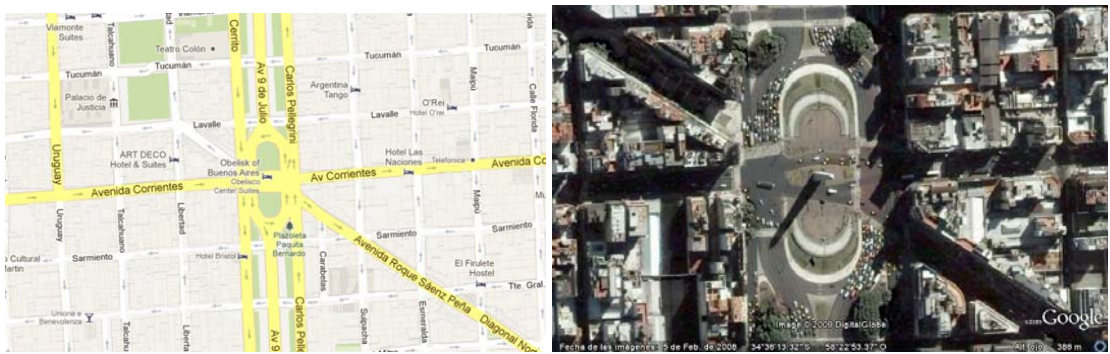

**Figura 11: Mapa y vista aérea de la zona del Obelisco en la parte céntrica de la ciudad de Buenos Aires. La avenida 9 de Julio corre verticalmente en ambas figuras, aproximadamente de norte a sur (el norte está arriba de la hoja). La sombra del Obelisco, bien visible en la imagen de la derecha, nos indica que la foto fue obtenida durante la mañana, unas pocas horas antes del mediodía solar.**

Dicho interrogante, supone un primer momento de intercambio de ideas entre docentes y alumnos, y permite indagar los saberes previos de estos últimos. Algunas preguntas útiles para plantear durante la discusión podrían ser las siguientes:

- ¿A qué se debe que la dirección de las sombras cambie y lo mismo su tamaño? ¿Por qué sucede esto? (Discusión sobre la trayectoria aparente del Sol.)
- ¿Cuánto tiempo deberá transcurrir para que el obelisco (que actúa como nuestro gnomon) proyecte sobre el suelo dos sombras de aproximadamente la misma longitud? (Buscamos obtener mayor precisión sobre la forma del arco diurno solar.)
- ¿Cómo se llama el momento del día que separa aproximadamente las sombras de un lado y del otro de la avenida 9 de julio? ¿Con qué hora civil se corresponde? (Discusión relativa a horario civil *versus* horario solar.)
- ¿Qué longitud tendrá la sombra en ese momento, comparada con su longitud en otros momentos del día? ¿Qué implica esto con relación a la altura del Sol sobre el horizonte?

• ¿Existe, entonces, alguna manera de ubicarnos espacialmente (y de orientarnos) haciendo uso de nuestro obelisco? ¿Cómo lo haríamos? ¿Cuánto tiempo nos llevaría?

Las preguntas anteriores nos llevan a poner el foco de la discusión en el recorrido aparente del Sol y en cómo cambia la sombra de un objeto cualquiera con el correr del tiempo. El intercambio también nos permite descubrir que, independientemente de la época del año, en cada día siempre hay una sombra que es la más corta de todas, y que se orienta a lo largo de la meridiana local del sitio de observación. En el caso de sitios ubicados por fuera de la zona tropical del planeta, esa sombra apuntará hacia el norte (en el hemisferio norte) o hacia el sur (en el hemisferio austral; este es precisamente el caso de Buenos Aires). Por el contrario, en el caso de sitios ubicados en la franja intertropical, la orientación norte o sur dependerá de la época del año en la que se trabaje. En cada uno de esos sitios habrá un día (en realidad, hay dos de esos días) en el que la sombra más corta tiene longitud nula (pues el Sol se halla en el cenit, al mediodía solar). Un par de días antes y un par de días después de ese día, las sombras correspondientes del mediodía solar apuntarán en sentidos contrarios.

Luego de esta breve "entrada en tema", a continuación debemos pasar de un gnomon real, como el obelisco, ubicado en un sitio particular de la ciudad, a un gnomon simulado, ubicado en el lugar apropiado de nuestro globo terráqueo paralelo. Para hacerlo, podríamos simplemente colocar un palillo vertical sobre el GTP, precisamente en la ciudad en donde nos hallamos (si hemos orientado el GTP adecuadamente, esa ciudad se ubicará arriba de todo). El palillo lo podríamos fijar con plastilina o masilla para que quede rígido y no se mueva durante nuestro trabajo (si se mueve, su sombra también lo hará, y confundirá nuestras observaciones). Pero, como lo mencionamos antes, un objeto del tamaño adecuado para trabajar resultará desmesuradamente fuera de escala: un palillo de 2 cm de longitud sobre un globo de 15 cm de radio representa un obelisco de más de 850 km de altura. Y por supuesto, por el momento no hay nada, ni obelisco ni montaña, que alcance esas dimensiones sobre nuestro planeta.

¿Deberemos entonces contentarnos con un palillo-gnomon de tamaño irrealmente grande en nuestro trabajo didáctico con el GTP? Creemos que existe una solución interesante para recuperar la escala y las dimensiones adecuadas en nuestro trabajo de modelización. El docente puede sugerir a sus alumnos que construyan una "maqueta" simplificada del obelisco y de sus zonas aledañas, todo en dimensiones pequeñas, pero respetando las proporciones (ESTEBAN, 2009). Una vez hecho esto, puede colocarse la maqueta en el lugar apropiado del GTP, con la orientación adecuada, y tendremos resuelto el problema. Dispondremos así de un "simulador" de las sombras del obelisco que representa fielmente no solo la orientación de las sombras sino también sus dimensiones (Figura 12).

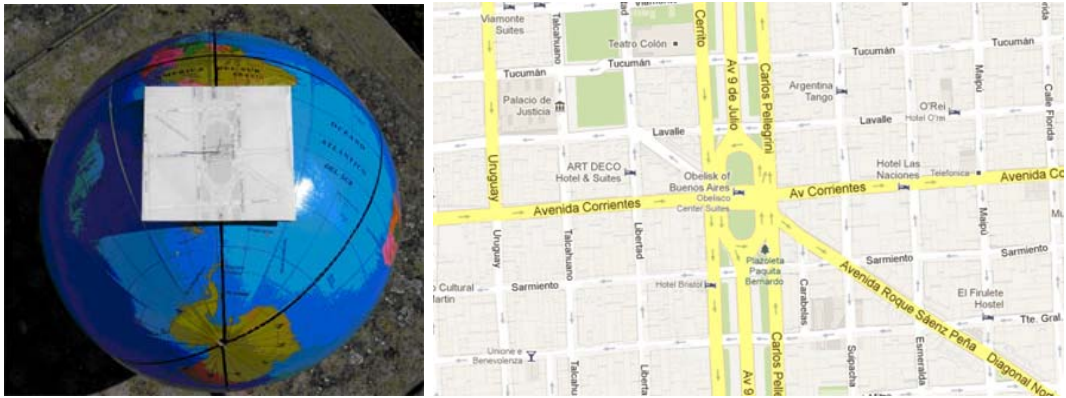

**Figura 12: En la foto de la izquierda se ve el dispositivo que permite estudiar las sombras proyectadas por un obelisco en el trabajo con el GTP. En este caso se trata del obelisco de la ciudad de Buenos Aires. El mapa cuadrado, pegado sobre un cartón y apoyado sobre el globo es un recorte del mapa de los aledaños del monumento y el alambre vertical que representa al obelisco tiene las dimensiones adecuadas, en proporción con el mapa (el monumento mide unos 67m de altura). En la foto se ve el polo sur del GTP y el mapa está orientado de manera que el norte quede hacia arriba de la hoja (la misma orientación del mapa de la derecha, que ya apareció en una figura anterior). Haciendo el seguimiento de la sombra del obelisco, es posible detectar el momento aproximado en el cual ésta cruza de un lado al otro de la avenida 9 de julio (que corre aproximadamente de sur a norte). Como en Buenos Aires el Sol al mediodía siempre se halla del lado norte, la sombra más corta del obelisco indicará la dirección hacia el sur (hacia abajo en la imagen).**

Dado que el GTP ha sido adecuadamente orientado y dado que la maqueta del obelisco y sus aledaños fue fabricada respetando las escalas y ubicada en el lugar correcto, las sombras que veamos sobre esta última son representación fiel y precisa (dentro de nuestra aproximación pedagógica) de las sombras verdaderas que el monumento proyecta en las calles y plazas de esa parte de la ciudad. En resumen, con nuestro dispositivo y sin necesidad de desplazarnos hacia ese lugar tan concurrido del microcentro de Buenos Aires, podemos saber con precisión cómo son las sombras y responder así la pregunta que inicialmente motivó nuestra actividad.

## 8. Obeliscos por todo el planeta

El dispositivo con el que venimos trabajando todavía puede brindarnos más servicios (y sorpresas). ¿Por qué limitarnos a conocer tan solo las sombras de nuestra ciudad? Veremos aquí que el GTP nos permite conocer la evolución de las sombras de cualquier objeto que se halle en cualquier parte de la Tierra, con la única restricción de que debe ser un sitio que, como el nuestro, esté iluminado por la luz del Sol (en otras palabras, que el terminador no lo haya dejado en las sombras).

Comencemos primero con algo cercano. Con el GTP adecuadamente alineado, coloquemos un simple palillo encima de nuestra ciudad. Este deberá estar orientado radialmente, es decir, en la dirección que apunta al centro geométrico del globo. Al lado (a unos metros) del GTP coloquemos un palo de madera de aproximadamente 1 metro de longitud, parado, haciendo las veces de un gnomon recto vertical. Dejando pasar las horas, veremos que las sombras evolucionan de idéntica manera en ambos gnomones. Es decir, en ambos casos, por ejemplo, las sombras cruzan el meridiano local a aproximadamente la misma hora. Esto nos asegura que el dispositivo funciona bien.

Pasemos ahora a lo que anunciamos unas líneas más atrás: distribuyamos palillos de 2 cm por todos lados en el GTP.

Para ir en orden, ubiquemos el meridiano que pasa por nuestra ciudad y aquellos meridianos cercanos a este. Coloquemos palillos "radiales" distribuidos por la superficie del globo, pero en sitios con coordenadas tales que la longitud sea constante y las latitudes varíen. En otras palabras, ubicaremos palillos a lo largo de los meridianos del GTP, como se muestra en la Figura 13. Para hacer la situación lo más simple posible, consideremos un equinoccio, que es el único momento del año en que el Sol llega a iluminar toda la Tierra y se halla equidistante de los polos. Durante esos dos días del año, como vimos, el Sol viaja a lo largo del ecuador celeste y, por ello, los sitios con latitud cero (círculo ecuatorial) tendrán al Sol en el cenit en su mediodía solar y los objetos allí ubicados solo proyectarán sombras hacia el oeste o hacia el este.

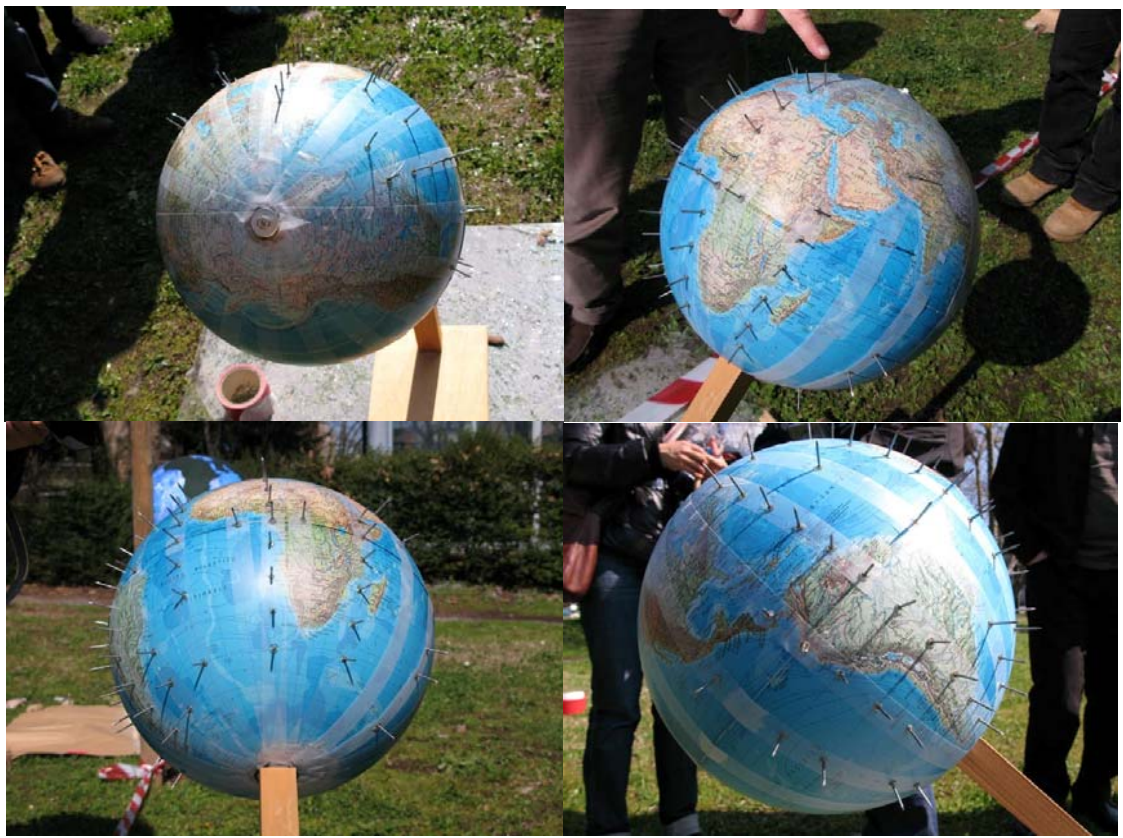

**Figura 13: Cuatro vistas de un GTP ubicado en cercanías de Milán, Italia, a la hora del mediodía solar del equinoccio de marzo (del año 2011). Las fotografías fueron tomadas desde el norte (primera foto), desde el este (muestra África y el dedo señala la ciudad de Milán), desde el sur (con el palo incrustado en el polo sur) y desde el oeste (muestra Sudamérica). Los palillos radiales alineados a lo largo de los meridianos permiten visualizar muy bien la diferente orientación de las sombras en este horario. Nótese que dado que estamos en un equinoccio, la trayectoria aparente del Sol en el cielo coincide con el ecuador celeste. En consecuencia los palillos-gnomones ubicados en el círculo ecuatorial solo proyectan sombras a lo largo de dicho círculo. Imágenes cortesía del grupo de Sesto San Giovanni y del proyecto www.globolocal.net.**

De estas imágenes podemos verificar que las sombras aumentan su longitud a medida que nos alejamos del ecuador, en ambas direcciones, y son tanto más largas – para una hora solar fija– cuanto mayor es la latitud. Y no solo eso: nótese además que cuanto mayor es la latitud del sitio, tanto más se inclina la dirección de la sombra con respecto a aquella de la línea ecuatorial, exceptuando, por supuesto, a las sombras de los gnomones ubicados en el meridiano donde es, justamente, el mediodía solar. A lo largo de este último meridiano, la hora solar es exactamente la misma y los palillos proyectan sus sombras en dirección norte o sur (Figura 14).

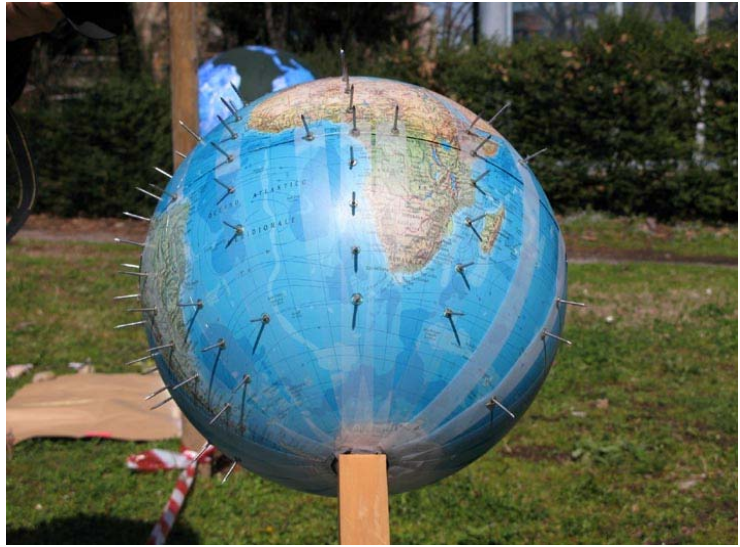

**Figura 14: Vista desde el sur del GTP de la figura anterior, ubicado en coordenadas (45º32'N, 09º14'E), hacia el noreste de Milán, en el mediodía solar de un equinoccio. Nótese que los palillos ubicados sobre el círculo ecuatorial solo proyectan sombras sobre ese círculo y que la desviación de las sombras respecto a la dirección este-oeste es tanto mayor cuanto más nos alejamos del ecuador. Entre otras varias consecuencias, esto último muestra claramente que los relojes solares horizontales deben ser construidos teniendo en cuenta la latitud del sitio donde serán usados (GANGUI, 2012). Dado que sitios sobre un mismo meridiano tienen la misma hora solar, la diferencia en las inclinaciones de las sombras que se muestran en esta imagen nos indica que los cuadrantes de estos relojes cambian si se cambia la latitud (ROHR, 1986).**

A continuación, de manera similar a lo ya hecho, podemos colocar palillos pero ahora no distribuidos a lo largo de los meridianos sino a lo largo de los paralelos del GTP. Como estas son circunferencias cuyos planos se orientan perpendicularmente al eje de rotación del planeta, todos los palillos de un mismo paralelo "harán lo mismo", cada uno en su momento. Es decir, las sombras que proyecte un palillo con el correr de las horas serán las mismas que aquellas que proyecten sus vecinos del mismo paralelo (horas antes u horas después).

A diferencia de lo que sucede con la latitud de un observador, que si conoce la época del año, las sombras le indicarán fielmente a qué distancia se halla del ecuador, hallar la longitud geográfica mediante las sombras no es una tarea simple. Por ejemplo, las sombras de un gnomon ubicado en Ciudad del Cabo (a una latitud de casi 34° S) serán prácticamente las mismas que las de un gnomon ubicado en la ciudad de Montevideo (latitud de casi 35° S) unas cinco horas más tarde.

En consecuencia, las sombras por sí solas nos dicen muy poco sobre nuestra ubicación "en longitud" sobre la superficie terrestre. El lector quizás recordará que este problema fue uno de los que aquejó durante muchos años a los marinos de ultramar, quienes no contaban con ningún método astronómico simple (léase, usando las sombras) para saber en qué meridiano de la Tierra (o del océano) se hallaban (SOBEL, 1995).

Continuando con nuestra secuencia de actividades, hay muchos otros temas que podemos discutir con alumnos de los últimos años de la escolaridad secundaria y de formación docente. Por ejemplo: si imaginamos que la sombra de un gnomon se asemeja a la manecilla principal de un reloj analógico, ¿para qué lado gira esa manecilla? A menos que elijamos muy sutilmente el sitio sobre el GTP, cualquier día del año, en prácticamente cualquier lugar de la Tierra, la sombra de un objeto tendrá un sentido de giro muy bien definido (a favor o en contra de las agujas del reloj). Sitios ubicados hacia el norte del Trópico de Cáncer y sitios ubicados hacia el sur del Trópico de Capricornio tendrán siempre sentidos de giro (de las sombras) opuestos. Esta es una constatación que puede hacerse simplemente con el dispositivo que venimos usando. Nótese que en la "zona tórrida", entre los trópicos, el Sol puede pasar por el lado sur o por el lado norte del cenit de un observador, dependiendo de la época del año en la que se halle. Esto hará que el sentido de giro de las sombras también cambie.

Para finalizar estas actividades, veamos algunos tópicos o preguntas adicionales que podemos plantear durante el desarrollo de la propuesta. Por ejemplo:

- Analicemos en qué lugares del planeta se da la situación que en algún día del año, al mediodía solar, un gnomon vertical no da sombra. (El docente puede sugerir emplear varios palillos ubicados en los distintos paralelos pero de un único meridiano del GTP.)
- ¿Tiene esto último alguna relación con la existencia de las líneas de los trópicos? (Esto justifica la presencia de esos dos paralelos singulares de la Tierra.)
- Si colocamos un palillo justo en el polo sur (o en el polo norte, dependiendo del hemisferio de residencia del observador), ¿cómo serán las sombras de este gnomon a lo largo del día? (Por la latitud extrema del sitio, es de suponer que los resultados serán sorprendentes y muy distintos de acuerdo a la época del año en la que se haga la observación.)
- Un dispositivo tan simple como el GTP, ¿permitirá estimar la hora local en cualquier lugar iluminado del planeta?
- Discutamos si existen zonas del GTP que merezcan el apodo de "países de las sombras largas". (En cualquier día del año y en cualquier momento del día, el GTP nos permitirá visualizar las sombras de las zonas polares y verificar esta afirmación.)

Quizás también sea útil para estas actividades ofrecer a nuestros alumnos la posibilidad de confeccionar un cuadro con los datos que a ellos les resulten destacables. Los mismos alumnos podrán sugerir los datos a recabar de sus observaciones con el GTP, pero en caso de que el docente lo crea necesario, aquí ofrecemos un posible cuadro modelo de registro donde los alumnos podrán volcar sus observaciones para luego analizar e interpretar los datos entre todos (Cuadro 1).

| Ubicación del gnomon | Hora local | Descripción de la sombra | Comentarios |
|---|---|---|---|
| | | | |

**Cuadro 1: Modelo de cuadro de registro de datos para el trabajo de los alumnos con el globo terráqueo paralelo.**

## 9. Consideraciones finales

La astronomía y la cosmografía están poco representadas en la educación formal en nuestro país, relegándose su dictado, en muchos casos, a proyectos especiales sostenidos desde la buena voluntad de los docentes. La responsabilidad de estos temas recae generalmente en profesores de física, matemática o geografía, para quienes, durante su formación, la astronomía no tuvo un lugar significativo.

Por otra parte, el análisis crítico de los diseños curriculares para la enseñanza primaria y niveles superiores permite corroborar la importancia otorgada a temas relacionados con los fenómenos astronómicos. Este es el caso en muchas de las jurisdicciones de nuestro país. Además, es reconocido el interés que "el espacio" genera en los alumnos, lo que brinda una excelente oportunidad para que los docentes trabajen algunas ideas que resultan ser muy interesantes y atractivas, aunque también problemáticas.

La existencia de obstáculos e ideas previas ya largamente estudiados en alumnos y docentes (CAMINO, 1995; POZO, 1999; MARTÍNEZ SEBASTIÀ, 2004; GANGUI et al., 2010), y la falta de reflexión sobre los diversos temas astronómicos presentes en los diseños curriculares, vuelven necesario el trabajo de aula con materiales concretos, la modelización y el empleo de recursos novedosos. Se hace inminente una didáctica de la astronomía que contribuya, entre otras cuestiones, con recursos que aporten reales herramientas de trabajo en las clases.

El dispositivo aquí propuesto, aunque ya conocido y empleado por algunos investigadores desde hace años (LANCIANO et al., 2011), aun no ha alcanzado el lugar que merece en la planificación usual de las clases de astronomía, quizás por falta de secuencias de trabajo guiadas y sencillas, recursos útiles para que los docentes puedan volcarlos en sus clases. Las actividades que aquí hemos propuesto intentan ayudar a paliar esta situación y a favorecer el empleo del globo terráqueo paralelo para reflexionar sobre la "astronomía diurna" (e.g., CAMINO y GANGUI, 2012), ya sea entre alumnos de la escolaridad primaria y secundaria, como entre los docentes en formación. De esta manera, intentamos colaborar para fomentar el desarrollo de aprendizajes significativos en el área de la astronomía.

## Agradecimientos

## Referencias